\documentclass[conference]{IEEEtran}
\IEEEoverridecommandlockouts
\usepackage{color, soul}
\usepackage{float}

\usepackage{cite}
\usepackage{amsmath,amssymb,amsfonts}
\usepackage{algorithmic}
\usepackage{graphicx}
\usepackage{textcomp}
\usepackage{xcolor}
\usepackage{braket}
\usepackage{balance}
\usepackage{setspace}
\usepackage{amsmath}

\renewcommand{\baselinestretch}{0.964}
\begin{document}

\title{
Optimal Mapping for Near-Term Quantum Architectures based on Rydberg Atoms
}

\author{
\thanks{This work was partially funded by the Carl Zeiss foundation.}

\begin{tabular}{@{}c@{\quad}c@{}}
\multicolumn{2}{c}{Sebastian Brandhofer$^{1}$ \qquad Hans Peter Büchler$^2$ \qquad Ilia Polian$^{1}$}\\[0.5cm]
    \begin{tabular}{@{}c@{}}
\normalsize
    $^1$Institute of Computer Architecture and Computer Engineering\\
\normalsize
    Center for Integrated Quantum Science and Technology\\
\normalsize
    University of Stuttgart, Stuttgart, Germany \\
\normalsize
    \{sebastian.brandhofer\;$|$\;ilia.polian\}@iti.uni-stuttgart.de
    \end{tabular}
&
    \begin{tabular}{@{}c@{}}
\normalsize
    $^2$Institute for Theoretical Physics III \\
\normalsize
    Center for Integrated Quantum Science and Technology\\
\normalsize
    University of Stuttgart, Stuttgart, Germany \\
\normalsize
    buechler@theo3.physik.uni-stuttgart.de\hspace{0cm}\\

    \end{tabular}

\end{tabular}
\vspace{-1ex}
}

\maketitle

\begin{abstract}
Quantum algorithms promise quadratic or exponential speedups for applications in cryptography, chemistry and material sciences.
The topologies of today's quantum computers offer limited connectivity, leading to significant overheads for implementing such quantum algorithms.
One-dimensional topology displacements that remedy these limits have been recently demonstrated for architectures based on Rydberg atoms, and they are possible in principle in photonic and ion trap architectures.
We present the first optimal quantum circuit-to-architecture mapping algorithm that exploits such one-dimensional topology displacements. We benchmark our method on quantum circuits with up to 15 qubits and investigate the improvements compared with conventional mapping based on inserting swap gates into the quantum circuits.
Depending on underlying technology parameters, our approach can decrease the quantum circuit depth by up to 58\% and increase the fidelity by up to 29\%.
We also study runtime and fidelity requirements on one-dimensional displacements and swap gates to derive conditions under which one-dimensional topology displacements provide benefits.

\end{abstract}

\vspace{-2ex}
\section{Introduction}\label{sec:intro}
The availability of ever-increasing quantum resources in contemporary quantum computers promises disruptive applications in drug design~\cite{drugdesign}, chemistry~\cite{microsoft_nitrogen_fixation}, material sciences~\cite{materialscience}~and cryptography~\cite{Shor}. 
However, the computation of quantum algorithms facilitating advances in these fields is limited by large error rates and short decoherence times in current quantum computers~\cite{Preskill2018, arsonisq}.
Furthermore, in most quantum computing technologies such as ion traps~\cite{ionq,ion2,quantum_inspire, honeywell}, superconducting circuits~\cite{GoogleSupremacy, ibm, rigetti}, Rydberg atoms~\cite{rydberg_pascal}~and NV centers~\cite{nv2}~only a subset of qubit-qubit interactions defined by the \emph{topology} of a quantum computer are supported at a time.
Quantum algorithms can be made compliant with a quantum computer topology through \emph{quantum circuit mapping}.
Quantum circuit mapping, however, incurs further errors during quantum algorithm computations since additional operations, e.g. swap gates, need to be inserted into the computation~\cite{swap_overhead, mapping_quantum_teleportation}.
It is therefore crucial to reduce the overhead incurred by quantum circuit mapping to extend the set of feasible quantum algorithm computations in current noisy and intermediate-scale quantum computing technology~\cite{Preskill2018}.

Quantum computing architectures based on Rydberg atoms are characterized by long decoherence times, high-fidelity quantum gates~\cite{rydberg_high_fidelity_two}, multi-qubit interactions~\cite{rydberg_multi_qubit_gate}~and the capability to dynamically change their topology during the computation of a quantum algorithm~\cite{17, Barredo2016,rydberg_optical_tweezers2}.
They are therefore a promising candidate for near-term quantum computing and physical realizations are currently investigated by research groups~\cite{QRydDemo}~and startups such as QUERA~\cite{QUERA}~and~Pasqal~\cite{Pasqal}.
However, the topology changes supported by quantum computers based on Rydberg atoms have not been explored for quantum circuit mapping, yet.

Topology changes, which do not manipulate the quantum state \emph{per se}, can have lower error rates than swap gates. Moreover, one topology change can have the same effect as multiple swap gates, potentially reducing the circuit's depth and increasing the probability to complete the circuit within the limit given by the decoherence time.
The computational reach of quantum technologies supporting topology changes, such as Rydberg atoms, may therefore be extended.

In this work, we develop an optimal method that exploits the topology changes available in Rydberg atom quantum computing technologies for quantum circuit mapping by:
\begin{itemize}
    \item developing a novel formal model for one-dimensional topology changes and extending it with a given formal model for swap gate insertion.
    \item evaluating the developed quantum circuit mapping method on intermediate-scale quantum circuits.
    \item characterizing technology parameters that enable an improved quantum circuit mapping through topology changes supported by near-term Rydberg architectures.
\end{itemize}
The remainder of this work is structured as follows. In section~\ref{sec:qc}~the principles of quantum computing are introduced.
Section~\ref{sec:ryd}~introduces quantum architectures based on Rydberg atoms and topology changes supported by near-term devices.
Section~\ref{sec:qc_mapping}~introduces quantum circuit mapping and reviews related work. 
Section~\ref{sec:method}~introduces a novel formal model that considers topology changes supported by the Rydberg atoms platform and discusses how to use this model for conventional quantum circuit mapping using swap gates.
Section~\ref{sec:results}~shows the improvement in quantum circuit depth and fidelity by considering such topology changes for different technology parameters on intermediate-scale quantum circuits.
Section~\ref{sec:conclusion}~concludes the work.
\section{Quantum Computing}\label{sec:qc}
A quantum computer performs computations on the quantum state of $n$-qubits as specified by external control.
The basic unit of information in quantum computing is the qubit, which describes a
two-level quantum system~\cite{qc10th}.
The quantum state of one qubit can be described as
\begin{equation}
\ket{\psi} = \alpha_{0}\ket{0}+\alpha_{1}\ket{1},
\end{equation}
where $\ket{i}$ are the computational basis states with $\alpha_{i}$ complex probability amplitudes whose magnitudes must sum up to one.
If more than one complex probability amplitude is larger than zero, the state is said to be in a superposition.
The measurement of a state $\ket{\psi}$ yields the result $i$ with probability $|\alpha_{i}|^2$.
If the result $i$ was measured, the state typically collapses to the computational basis state $\ket{i}$.
Quantum gates describe quantum state transformations by defining changes to the complex probability amplitudes of the state.
A quantum computer typically supports a subset of quantum gates, e.g. IBM's quantum computers currently support quantum gates for specific single-qubit rotations and the controlled-not (CX) two-qubit quantum gate.
Furthermore, a quantum computer may only support multi-qubit gates between specific qubits~\cite{GoogleSupremacy, ion2, k_brown}.
Such qubit-qubit interaction constraints are described by the \emph{topology graph} of a quantum computer.

\begin{figure}[t!]
  \centering
  \includegraphics[width=0.8\linewidth]{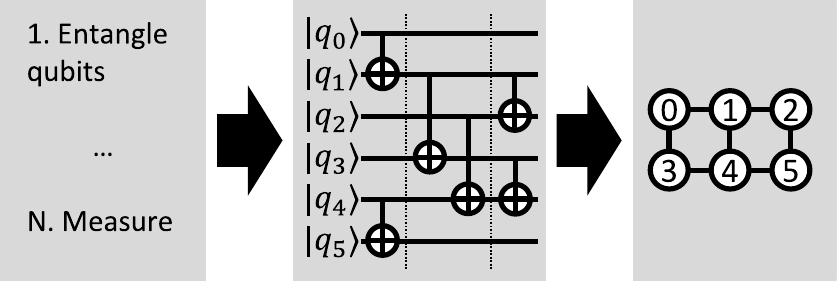}
  \caption{Synthesis of the first steps of a quantum algorithm (left) to a six-qubit quantum circuit (middle) that is mapped to the topology graph (right) of a six-qubit quantum computer.\label{fig:mapping}}
  \vspace{-2ex}
\end{figure}

Figure~\ref{fig:mapping}~depicts the steps required to perform a quantum algorithm on a quantum computer.
A quantum algorithm describes how to transform a quantum state such that the solution to a computational problem is yielded.
Before executing a quantum algorithm on a target quantum computer, the algorithmic description must first be compiled to a quantum circuit that contains supported quantum gates and is then mapped to the topology of the target quantum computer.
We call the latter compilation step \emph{quantum circuit mapping} and the former \emph{quantum algorithm synthesis}. This paper focuses on the mapping step.

Executing a quantum algorithm poses several requirements on the target quantum computer. 
The target quantum computer must have at least as many qubits as required by the quantum algorithm.
Furthermore, the decoherence time of the quantum computer must be much larger than the minimum time a quantum algorithm computation requires on the quantum computer.
A quantum algorithm also poses requirements on the physical error rates of the employed quantum computer~\cite{arsonisq, vts_special_session}.

\section{Quantum Computing Based on Rydberg Atoms}\label{sec:ryd}

Quantum architectures based on Rydberg atoms consist of an array of individually trapped neutral atoms.
The trapping is achieved by optical tweezers that allow for the deterministic loading of the array by physically moving atoms~\cite{17,Barredo2016,rydberg_optical_tweezers2}.
Single-qubit gates are achieved by individual optical addressing of the atoms, while two-qubit gates are realized by exciting the atoms into a Rydberg state, where Van der Waals interaction is enhanced such that it mediates a strong interaction between neighboring atoms~\cite{first_rydberg_gate_theory,first_rydberg_gate_experiment,first_rydberg_gate_experiment2,rydberg_multi_qubit_gate,graham2019,rydberg_high_fidelity_two}.

While in principle one can envisage arbitrary movement of the atoms during a quantum computation, a major challenge is to avoid errors due to the dephasing of the qubit and the coupling of the qubit to motional degrees of freedom.
Therefore, the movement of atoms during the quantum computation is expected to be restricted in near-term architectures based on Rydberg atoms.
Restricting the movement of the atoms along a row of the two dimensional array with fixed ordering of the qubits can avoid such errors and is currently experimentally implemented in e.g.~\cite{QRydDemo}.
We call such movements \emph{one-dimensional displacements} and model them through changes to the topology graph of the quantum computer.

A topology graph represents the qubits and possible qubit-qubit interactions in a quantum computer. 
The topology graph $G=(P, E)$ contains one vertex for each qubit in the quantum computer and one edge $e=(u, v)$, if qubit $u$ and qubit $v$ in the quantum computer can interact with each other, i.e. participate in a multi-qubit gate.
A \emph{topology graph change} is a removal or an addition of at least one edge to the graph.

\begin{figure}[t!]
  \centering
  \includegraphics[width=0.6\linewidth]{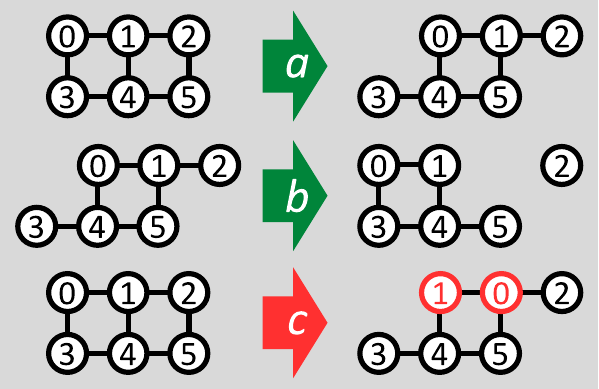}
  \caption{Potential topology graph changes in quantum computers based on Rydberg atoms. The topology graph changes \textit{a} and \textit{b} are valid one-dimensional topology displacements. The topology graph change \textit{c} is not a valid displacement since the relative qubit positions are not constant.
  \label{fig:topology_changes_example}}
  \vspace{-2.5ex}
\end{figure}

In this work, one-dimensional topology displacements are considered.
Figure~\ref{fig:topology_changes_example}~shows a $3\times 2$ grid where the rows are shifted relatively to each other in three different ways (\textit{a}, \textit{b} and \textit{c}).
In topology change \textit{a}, the qubits in the first row are displaced to the right side.
The second topology change (\textit{b}) displaces qubit 0 and qubit 1 to the left side by one and keeps qubit 2 in position.
It would also be possible to displace qubit 2 to the right side in the same topology change.
The topology change \textit{c} is not permitted since the relative position of qubit 0 and qubit 1 in the first row are swapped.

Depending on the physical realization of a quantum computer, a subset of topology graph changes are supported. 
Quantum computing technologies such as superconducting qubits and NV centers do not support deliberate changes in qubit-qubit interaction during the computation of a quantum algorithm and can therefore be represented by one static topology graph for quantum circuit mapping purposes~\cite{nv2,GoogleSupremacy}.
In contrast, quantum computing technologies based on photons, ion traps and Rydberg atoms can change the qubit-qubit interactions by physically moving qubits~\cite{wille_ion_trap, photon}~or changing other operational parameters~\cite{k_brown}.
In photonic quantum computers the qubit arrangement, and thus the qubit-qubit interaction, can be changed arbitrarily by placing mirrors~\cite{photon}~or waveguides~\cite{integrated_optics_topology_change}.
In ion trap quantum computer realizations, electromagnetic fields can be applied to support arbitrary topology changes~\cite{k_brown}.
Other ion trap quantum computers allow to physically move their ions depending on the placement and operation of electrodes in the quantum computer~\cite{wille_ion_trap}.

\section{Quantum Circuit Mapping}\label{sec:qc_mapping}

Quantum circuit mapping assigns each computation $c$ in a quantum circuit $U$ a \emph{location} $l\in L = \mathcal P(P)\setminus \{\}$ on a quantum computer to a set of qubits $P$ and a time step $t\in \{1, ..., T\}$~\cite{swap_gates, wille2014optimal}.
A computation $c$ may be a quantum gate, measurement or other operation and requires the specification of at least one qubit.
A location $l$ may be one or multiple qubits on a quantum computer. 
The primary objective of quantum circuit mapping is to adapt a quantum circuit to the topology $G=(P, E)$ of a quantum computer such that the qubit-qubit interaction requirements of each computation $c$ are satisfied, i.e. all multi-qubit gates are assigned to multiple vertices $p\in P$ that are connected according to $E$.
If two successive computations were specified on the same qubits in the quantum circuit but are assigned to different vertices during quantum circuit mapping, the mapping procedure must use operations such as swap gates~\cite{swap_gates}~or quantum teleportation~\cite{mapping_quantum_teleportation}.

Figure~\ref{fig:mapping}~shows the topology graph for a six-qubit quantum computer and a six-qubit quantum circuit containing six two-qubit gates.
Consider the two-qubit quantum gates CX(1, 3) and CX(2, 4) between qubit pair $1, 3$ and qubit pair $2, 4$.
They can not be performed directly by the quantum computer since its topology does not support these interactions.
However, the quantum circuit in figure~\ref{fig:mapping}~can be mapped by swapping the qubit state of $1$ with $4$ before and after computing the considered two-qubit quantum gates.

Besides the adaptation to the topology of a quantum computer, quantum circuit mapping may be performed subject to secondary objectives that typically are expected to reduce the incurred error during a quantum algorithm computation.
Such secondary objectives are typically the minimization of operations inserted by quantum circuit mapping and the minimization of the resulting quantum circuit depth~\cite{swap_gates,jku2,wille_minh, sabre, wille2014optimal, swap_overhead}.
However, further objectives have been proposed such as reduction of concurrent operations that incur crosstalk errors~\cite{crosstalk_mapping}, the maximization of circuit fidelity~\cite{variability_mapping, OLSQ}~or mapping operations to qubits that have demonstrated lower error rates during calibration protocols~\cite{variability_mapping, variability_mapping2}.

Heuristic~\cite{jku2, sabre, variability_mapping, variability_mapping2, crosstalk_mapping, swap_overhead}~and optimal~\cite{swap_gates, wille_minh, wille2014optimal}~algorithms for quantum circuit mapping have been proposed.
Prior quantum circuit mapping approaches relying on swap gate insertions consider topology graphs that remain constant during the computation of the quantum circuit~\cite{swap_gates,jku2,wille_minh, sabre, crosstalk_mapping, variability_mapping, variability_mapping2}.
Quantum circuit mapping methods specifically for ion trap quantum computers consider arbitrary qubit-qubit interactions through a sequence of topology graph changes~\cite{wille_ion_trap, ion_trap_mapping}.
However, these works do not consider swap gate insertions alongside topology changes.
In this work, swap gates and topology graph changes are both considered for quantum circuit mapping.

\section{Quantum Circuit Mapping for Near-Term Quantum Architectures based on Rydberg Atoms}\label{sec:method}

In this section we describe how a quantum circuit mapping can be computed by constructing and solving an satisfiability modulo theories (SMT) model $\mathcal M$. 
Model $\mathcal M$ is the union of a model $\mathcal R$ that describes valid topology changes in a near-term Rydberg architecture and a model $\mathcal S$ that describes the effect of swap gate insertions on the quantum circuit.

\begin{figure}[t!]
  \centering
  \includegraphics[width=1.0\linewidth]{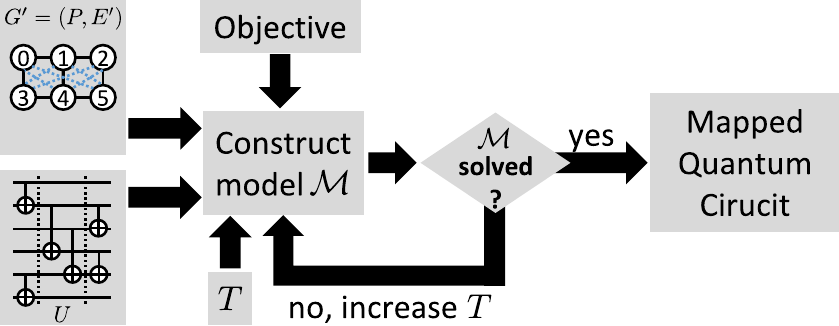}
  \caption{Individual steps of the developed quantum circuit mapping method for a quantum circuit $U$, topology graph $G'=(P,E')$ and maximal considered time steps $T$.\label{fig:flow}}
  \vspace{-3.5ex}
\end{figure}

Figure~\ref{fig:flow}~shows the steps of the developed quantum circuit mapping method for a quantum circuit $U$ and a topology graph $G=(P, E)$.
First, the topology graph $G$ is extended to yield a graph $G'=(P, E')$ that includes edges for all qubit-qubit interactions that can be supported by valid topology graph changes. 
The model $\mathcal M = \mathcal R \cup S$ is then constructed from the input quantum circuit $U$, topology graph $G'$ and the maximum quantum circuit depth $T$.
A solver then tries to determine valid assignments to $\mathcal M$ subject to further optimization objectives such as the minimization of the quantum circuit depth or the maximization of the circuit fidelity.

If the solver returns a valid assignment to $\mathcal M$, then the obtained quantum circuit mapping is optimal with respect to the chosen objective function.
Otherwise, model $\mathcal M$ is modified by increasing the maximum quantum circuit depth $T$ while retaining the original quantum circuit $U$ and extended topology graph $G'$.
This process is repeated until a valid assignment to $\mathcal M$, i.e. a quantum circuit mapping, is found.
Optimality is guaranteed with this approach, if the value of the employed objective function becomes worse for an increasing quantum circuit depth.
Otherwise, a sufficiently large initial maximum quantum circuit depth $T$ must be chosen that allows arbitrary qubit permutations and topology changes for each computation in the quantum circuit $U$.

The following sections assume that $P$ is the set of qubits accessible in a quantum computer, $Q$ are the qubits in a quantum circuit $U$ and $T$ is the maximum quantum circuit depth (or: last time step) considered for quantum circuit mapping.
$R$ $(C)$ is the number of rows (columns) in the topology and also a function that maps a qubit $p\in P$ to its corresponding row (column) in $G$.
Constraints that define the domain of a variable $v \in \mathcal R \cup \mathcal S$ have been omitted.

\vspace{-1ex}
\subsection{Topology Graph Extension}

\begin{figure}[t!]
  \centering
  \includegraphics[width=0.26\linewidth]{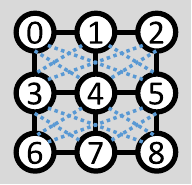}
  \caption{Extension of a topology graph by dashed, blue edges that may become available through one-dimensional topology displacements.\label{fig:topo_extension}}
  \vspace{-2ex}
\end{figure}

The first step is to construct the extended topology graph $G'=(V, E')$ according to topology displacements supported by near-term Rydberg atom architectures (see section~\ref{sec:ryd}).
We assume the Rydberg platform to initially have a topology $G$ where qubits are connected to their nearest neighbors and arranged in a $C\times R$ grid.
We model arbitrary displacements of qubits in one row along the x-axis of the grid such that the order of the qubits in the row is maintained.
The resulting extended topology graph $G'$ is shown in figure~\ref{fig:topo_extension}.
Since the qubits can be displaced along the x-axis of the grid, an arbitrary qubit $q$ of row $r$ can interact with an arbitrary qubit $u$ of a neighboring row $r'$.
Since the relative position of qubits in the same row may not change, qubits $q, q'$ in the same row can only interact if $\{q, 'q\}\in E$, where $E$ is the edge set of the original topology graph $G$.
Therefore, the edge set $E'$ of $G'$ is defined by
\begin{equation}
    E' = E \cup \{\{q, u\}\; | \; q\in r, u\in r'\}.
\end{equation}
The topology graph extension introduces $\mathcal O(C^2R)$ edges to $G$ for a $C \times R$ grid.

\subsection{Modeling One-Dimensional Topology Displacements}
The next step in the developed method is to define how a topology graph can change from one time step to another and which edges in $G'$ are available for computation in a specific topology displacement.
Given the extended topology graph $G'=(P, E')$ and the maximum considered quantum circuit depth $T$ the model of one-dimensional topology displacements (see section~\ref{sec:ryd}) $\mathcal R$ contains 
the following variables:
\begin{itemize}
    \item $\eta_{p, t}\in \mathbb{N} \quad \forall p\in P, \forall  t\in \{1,..., T\}$  --- represents the displacement of a qubit $p$ at time $t$, where a value of 0 indicates no displacement, a positive value indicates a displacement to the right side and a negative value indicates a displacement to the left side.
    \item $EN_{p, u, t}\in \mathbb{B} \quad \forall p, u\in P, \forall  t\in \{1,..., T\}$ --- indicates whether an edge in the extended topology graph $G'$ is available for a multi-qubit computation.
\end{itemize}
One-dimensional topology displacements are characterized by a fixed order of qubits in the same row for all time steps $t\in T$.
Each qubit $p\in P$ is assigned a displacement $\eta_{p, t}\in \mathbb{N}$ that is constrained by:
\begin{equation}
    \left(\eta_{p, t} \leq \eta_{r, t}\right) \wedge \left(\eta_{l, t} \leq \eta_{p, t}\right),
\end{equation}
where $r\in \{u \in P \;|\; \left(C(u) > C(p)\right) \wedge \left(R(u) = R(p)\right)\}$ and $l\in \{u \in P \;|\; \left(C(u) < C(p)\right) \wedge \left(R(u) = R(p)\right)\}$.

The displacements of qubit $p$ and qubit $u$ support an edge $e=(p, u)\in G'$ for a multi-qubit quantum gate computation at time $t$ if the displacements are equal
\begin{equation}
  EN_{p, u, t} := (\eta_{p, t} = \eta_{u, t})\\ 
\end{equation}
for qubits $p, u$ that are in the same row and
\begin{equation}
  EN_{p, u, t} := \left(C(p) + \eta_{p, t}\right) = \left(C(u) + \eta_{u, t}\right)
\end{equation}
for qubits $p, u$ that are in neighboring rows.

A topology displacement may require a runtime of $t_d \in \{1,..., T\}$ time steps.
A mismatch in displacement $\eta_{p, t} \neq \eta_{p, t+1}$ on a qubit $p\in P$ between time step $t$ and $t+1$ therefore implies that no other operation may be computed on $p$ in $t'\in \{t-t_d, ..., t\}$.
The runtime $t_d$ is a user-specified input to the optimization problem based on the experimentally measured duration of the displacement. 

\subsection{Formal Swap Gate Insertion Model}

Solving a swap gate insertion model $\mathcal S$ yields a quantum circuit mapping that inserts swap gates to adapt a quantum circuit $U$ to a topology graph $G$.
A swap gate insertion model defines an assignment of qubits $Q$ in a quantum circuit $U$ to the qubits of a quantum computer topology $P$ for each considered time step.
Furthermore, each computation $c \in U$ is assigned a location $l \in L = \mathcal P(P)\setminus \{\}$ for each time step.

In general, a swap insertion model $\mathcal S$ constraints the assignment to the following variables:
\begin{itemize}
    \item $\pi_{q, t} \in P  \quad \forall q\in Q, \forall  t\in \{1,..., T\}$ --- represents the assignment of a qubit $q$ in the input quantum circuit $U$ to a qubit $p\in P$ in the quantum computer.
    \item $l_{c} \in L \quad \forall c\in U$ --- indicates the location $l$ of a computation in the input quantum circuit $U$.
    \item $z_{c} \in \{1, ..., T\} \quad \forall c\in U$ --- indicates the time step $t$ of a computation in the input quantum circuit $U$.
\end{itemize}
under the following conditions:
\begin{itemize}
    \item Each computation $c\in U$ is assigned a location $l$ that satisfies the qubit requirement of $c$ and a time $z$ that satisfies the computation order defined in $U$.
    \item Each $q\in Q$ is assigned at most one $p\in P$, i.e. the assignment $\pi_{q, t}$ is unique at any fixed time step for all qubit $q\in Q$.
    \item If a computation $c\in U$ is assigned a location $P'=\{p_1, ...,p_n\}$ at time $z_c$, the qubits $Q'=\{q_0, ...,q_n\}$ specified by $c$ must have been assigned to $P'$, i.e $\pi_{q_i, z_c}=p_i\quad \forall q_i\in Q',p_i\in P'$.
    \item The assignment $\pi_{q, t}=p$ may only change to $\pi_{q, t+1}=p$ if there exists an edge $(p, u)$ in the topology graph $G$.
\end{itemize}
For each change in qubit assignment ($\pi_{u, t}\neq \pi_{u, t+1}=\pi_{q, t}\neq \pi_{q, t+1}$) a swap gate is inserted into the quantum circuit $U$.
Furthermore, operations may have individual runtimes $t_i\in \{1, ..., T\}$ that must be considered, i.e. a qubit $p\in P$ is occupied with at most one operation in any time step $t$. 

Combining a swap insertion model $\mathcal S$ with a model $\mathcal R$ for topology changes requires two more constraints:
\begin{itemize}
    \item A swap or multi-qubit computation may only act on pairs of qubits $p\in P, u\in P$ in a certain time step $t$ if $EN_{p, u, t}$ indicates that the edge between $p$ and $u$ is available in the topology at time $t$.
    \item Either a single-qubit gate, a multi-qubit gate, a topology displacement or a swap gate may act on a qubit at the same time step $t$.
\end{itemize}
Several formal swap gate insertion models were proposed in the state of the art~\cite{OLSQ, wille_minh, wille2014optimal}.
The swap insertion model $\mathcal S$ described in this section is generic and compatible with these approaches.
In our experiments reported in section~\ref{sec:results}~we will be using the specific model from~\cite{OLSQ}.

\subsection{Optimization Objectives}

Optimization objectives such as minimizing the number of inserted swap gates, maximizing the fidelity of the resulting quantum circuit or minimizing the resulting quantum circuit depth are crucial for obtaining a quantum circuit mapping that incurs low errors on the target quantum computer.

In model $\mathcal M$, the depth minimization and quantum circuit fidelity can be defined as:
\begin{itemize}
    \item minimize quantum circuit depth: $\min_{c\in U} z_c$.
    \item maximize quantum circuit fidelity: 
    \begin{equation*}
        \max \sum_{c\in U} \log(f_c) + \sum_{s\in S} \log(f_s) +\sum_{d\in D} \log(f_d),
    \end{equation*} where $f_c$ is the fidelity of a computation $c$, $S$ is the set of swap gates, $f_s$ is the fidelity of a swap gate $s\in S$, $D$ is the set of one-dimensional topology displacements and $f_d$ is the fidelity of a displacement $d\in D$.
\end{itemize}

\subsubsection*{\textbf{Example --- One-dimensional topology displacements improve quantum circuit mapping}}
We will now demonstrate model $\mathcal M$ on a quantum circuit that can be mapped with less overhead using one-dimensional topology displacements than using only swap gate insertions.
Consider the six-qubit quantum circuit and the $3\times 2$ topology grid $G$ in figure~\ref{fig:mapping}.
The quantum circuit consists of six two-qubit gates out of which two quantum gates CX(1, 3) and CX(4, 2) can not be computed on the specified topology graph $G$ directly.
Through the developed model $\mathcal M$, a quantum circuit mapping based on one-dimensional topology displacements or swap gate insertion can be computed.

Let the runtime of the swap gate and the topology displacement be $t_d$ and the runtime of all other operations be $t_i=1$.
The two two-qubit gates to the left of the dashed line in figure~\ref{fig:mapping}~can be computed directly in the first time step of the quantum computation.
\begin{figure}[t!]
  \centering
  \includegraphics[width=\linewidth]{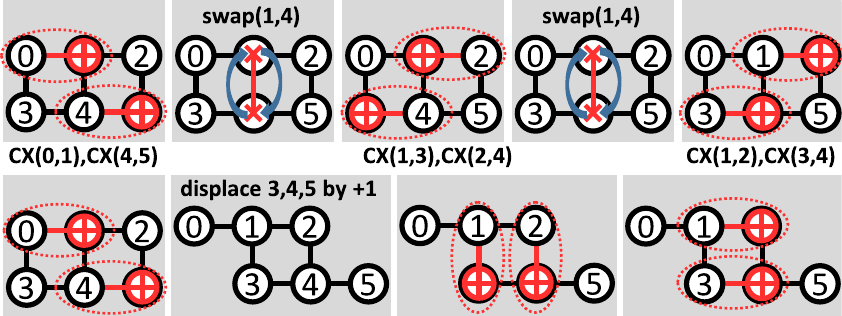}
  \caption{Mapping of the quantum circuit to the topology graph in figure~\ref{fig:mapping}.
  The top row shows the mapping with swap gate insertions only and the bottom row shows the mapping with an one-dimensional topology displacement.
  The row in-between shows the two-qubit gates that must be mapped.
  \label{fig:map_example}}
  \vspace{-3.5ex}
\end{figure}
In the case of a quantum circuit mapping using swap gate insertions, the quantum gates CX(1, 3) and CX(4, 2) can be computed in time step $t_d+2$ by inserting either set of swap gates $s_0=\{\text{swap}(1, 4)\}$ or $s_1=\{\text{swap}(1, 0),\; \text{swap}(4, 5)\}$.
If $s_0$ is computed until time step $t_d+2$, the remaining two-qubit gates can not be computed directly and would require another insertion of swap $s_0$.
This would lead to a resulting quantum circuit depth of $3+2t_d$.
If $s_1$ is computed until time step $t_d+2$, CX(0, 1) requires another swap. In this case, the resulting quantum circuit depth is again $3+2t_d$.

A quantum circuit mapping based on topology displacements can displace qubits $3, 4, 5$ by $1$ ($\eta_{3, 2+t_d}=\eta_{4, 2+t_d}=\eta_{5, 2+t_d}=1$).
This displacement allows interactions between qubits $1, 3$ and qubits $2, 4$, i.e. CX(1, 3) and CX(4, 2) are assigned the newly available qubit-qubit interactions as locations and time steps $2+t_d$ $(z_3=z_4=2+t_d)$.
The remaining two-qubit gates can be computed in the next time step, i.e. the resulting quantum circuit has depth $3+t_d$.
This example shows that even for small-scale quantum circuits, their structure can be exploited through one-dimensional topology displacements to reduce the depth of the mapped quantum circuit.

\section{Evaluation}\label{sec:results}
In this evaluation, we investigated the improvement in quantum circuit depth and quantum circuit fidelity when introducing one-dimensional topology displacements to quantum circuit mapping in addition to swap gates.
The improvement was evaluated for different technology parameters such as the fidelity of swap gates, and the runtime of topology displacements and swap gates.

We evaluated the developed quantum circuit mapping method on quantum circuits with up to 15 qubits and a maximum depth of $76$.
The evaluated quantum circuits compute arithmetic functions~\cite{jku}, Bernstein-Vazirani (BV)~\cite{bv}~or the quantum Fourier transformation (QFT)~\cite{qft}.
In addition, quantum circuits with multiple layers of two-qubit CX gates between random pairs of qubits were generated and evaluated.
For each evaluated quantum circuit with qubits $Q$, the topology graph was chosen to be a $C \times R$ grid where $R$ is the number of rows and $C$ is the number of columns such that $|Q|\leq C\cdot R$ and $C, R$ minimal.

The developed SMT model with the swap gate insertion model from~\cite{OLSQ}~was solved using the Z3 solver~\cite{z3solver}.
The average runtime of the solver was roughly $20$ minutes.
\subsection{Quantum Circuit Depth}
\begin{figure}[t!]
  \centering
  \includegraphics[width=0.55\linewidth]{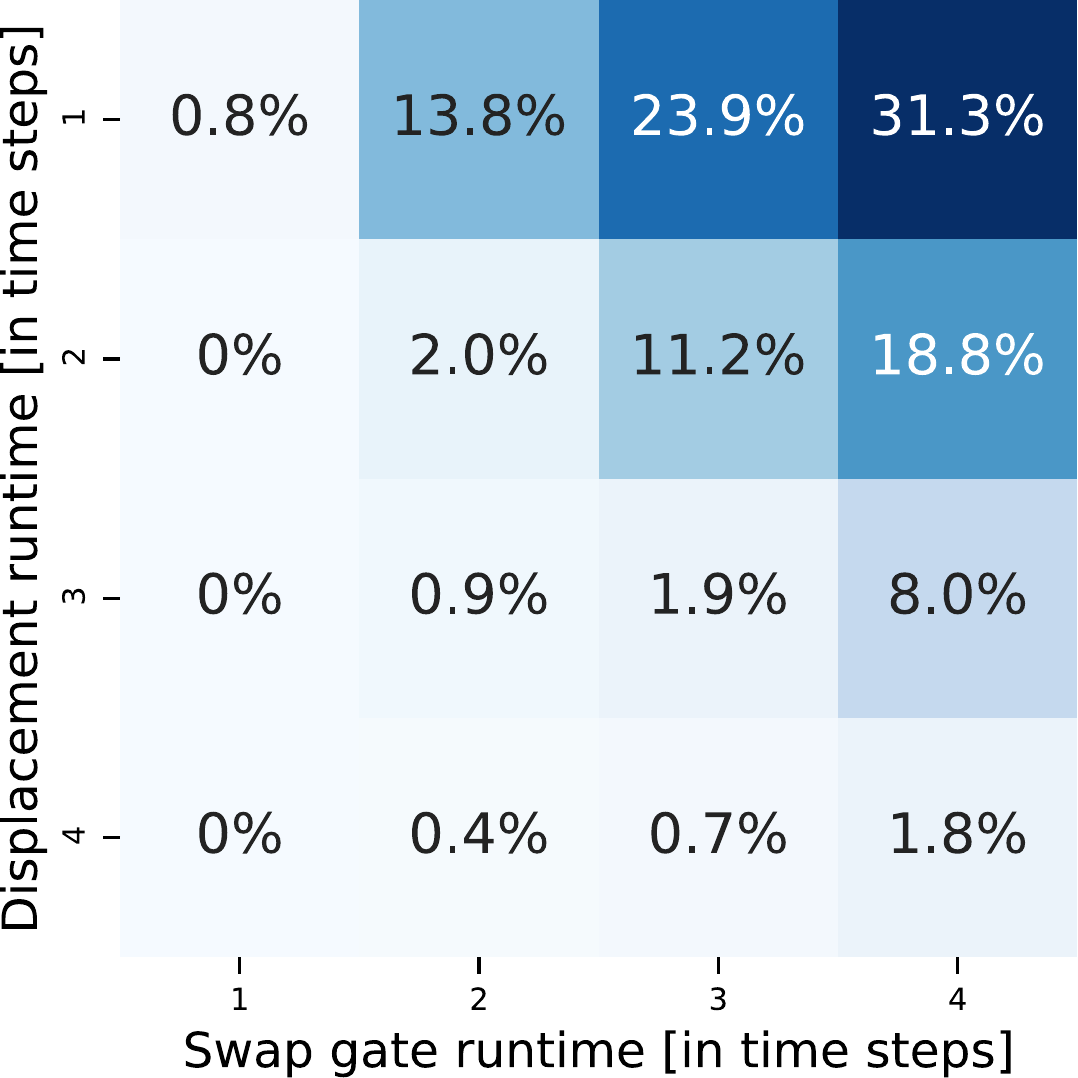}
  \caption{Quantum circuit depth reduction for different one-dimensional topology displacement and swap gate runtimes. \label{fig:heatmap}}
 \vspace{-3ex}
\end{figure}

Each quantum circuit was successively evaluated with a swap gate and topology displacement runtime ranging from one to four time steps.
Figure~\ref{fig:heatmap} shows the reduction in quantum circuit depth when using topology displacements in conjunction with swap gates for quantum circuit mapping.

If the runtime of swap gates and topology displacements is equal, the average reduction in quantum circuit depth ranges from $0.8\%$ to $2\%$ depending on the actual runtime.
If the swap runtime is smaller than the topology displacement runtime, the mapping procedure will typically fall back to inserting swap gates instead of using topology displacements unless one topology displacement can solve more qubit-qubit interaction requirements than one swap gate at a particular time step of the quantum circuit computation.
For the evaluated quantum circuits, this is never the case if the swap runtime equals one time step and the topology displacement runtime is larger than one time step.
However, if the swap runtime is two time steps and the topology displacement runtime is four time steps, a quantum circuit depth reduction of $0.4\%$ on average can be observed.

In contrast, if the swap runtime is larger than the topology displacement runtime, an average quantum circuit depth reduction of up to $31.3\%$ can be observed.
The exact quantum circuit depth improvement depends on the ratio $r$ between the swap gate runtime and topology displacement runtime.
If this ratio is $2$ the reduction in quantum circuit depth is $16.3\%$ on average.

As evident from the diagonal entries of the matrix in figure~\ref{fig:heatmap}, $rr$ is not the only factor that determines the quantum circuit depth reduction.
In the diagonal, the ratio of the two runtimes is always $1$, i.e. take the same number of time steps.
However, the depth reduction ranges from $0.8\%$ to $2\%$. 
This difference stems from the quantum circuit structure, i.e. the number and position of single- and multi-qubit gates.
\begin{figure}[t!]
  \centering
  \includegraphics[width=0.88\linewidth]{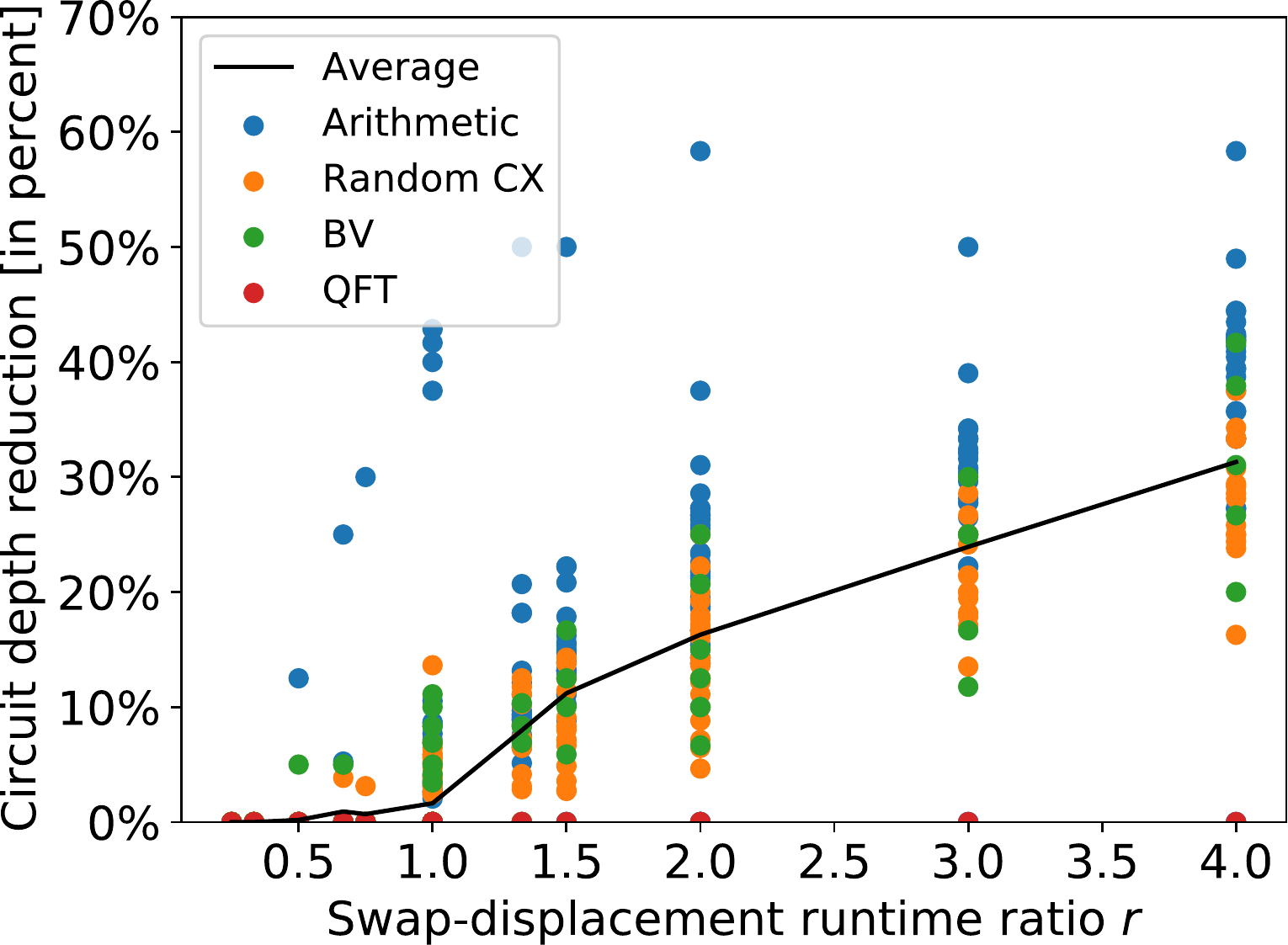}
  \caption{Quantum circuit depth reduction of individual quantum circuits for different runtime ratios $r$.\label{fig:depth_ratio}}
  \vspace{-1ex}
\end{figure}

The impact of the circuit structure is also evident in figure~\ref{fig:depth_ratio}.
The developed method does not reduce the depth of QFT quantum circuits at any evaluated combination of swap gate runtime and topology displacement runtime.
However, for arithmetic quantum circuits, the quantum circuit depth reduction is up to $58.3\%$.
If $r$ is $1$, the quantum circuit depth reduction is up to $45\%$.
With decreasing $r$, the structure of the quantum circuit becomes insignificant for the reduction of quantum circuit depth:
if the swap gate is twice as fast as the topology displacement, the maximum quantum circuit depth reduction is roughly $13\%$.
At even faster swap gates, the topology displacements do not have a tangible impact on the quantum circuit depth reduction.

\subsection{Fidelity}
\begin{figure}[t!]
  \centering
  \includegraphics[width=0.88\linewidth]{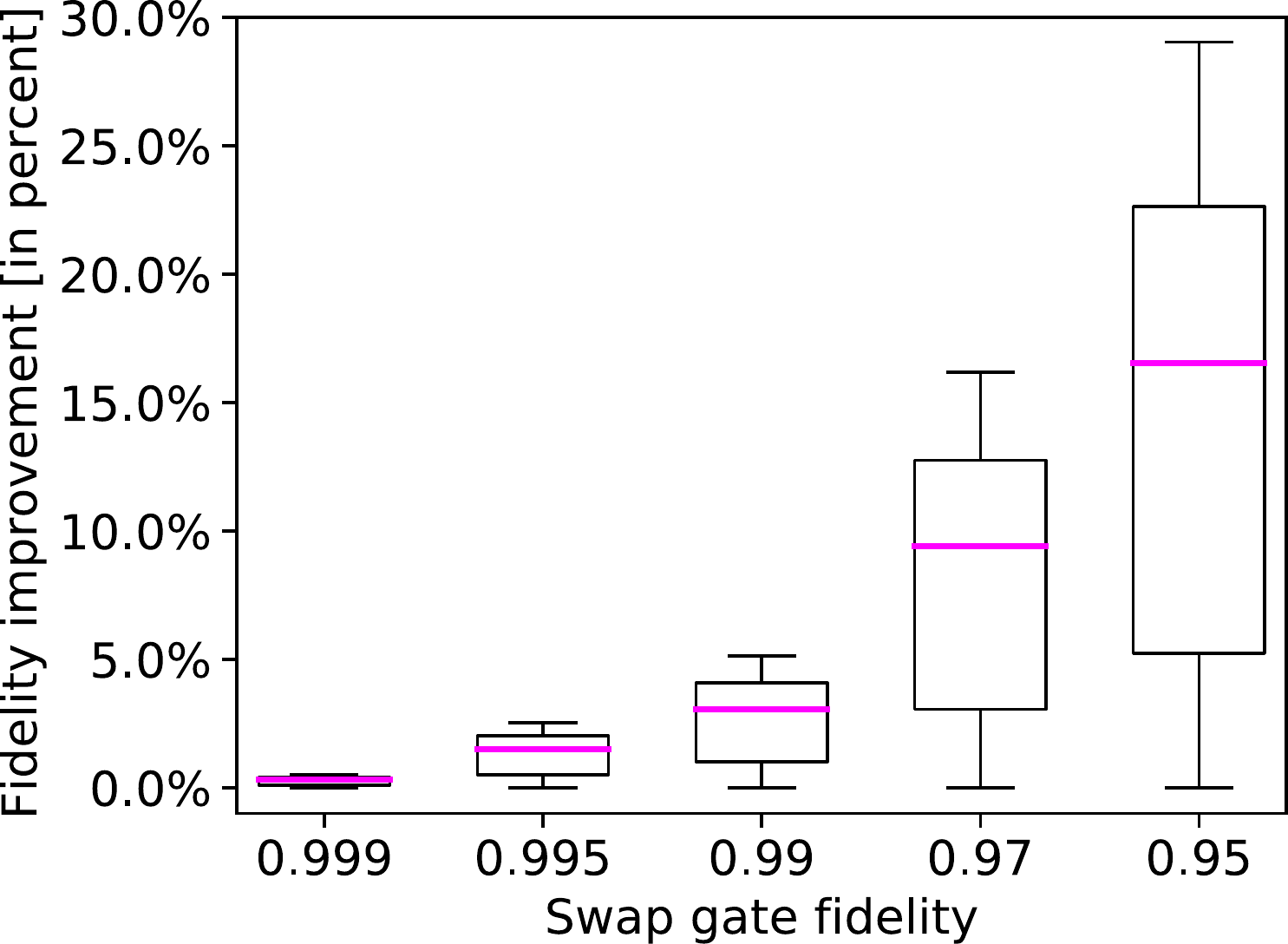}
  \caption{Quantum circuit fidelity improvement for different swap gate fidelities $f_s\in \{0.999, 0.995, 0.99, 0.97, 0.95\}$. \label{fig:fid_boxplots}}
 \vspace{-1ex}
\end{figure}

The fidelity was set to $1$ for topology displacements since we expect these restricted atom movements through optical tweezers to not have an impact on the quantum state of the qubits.
The evaluated quantum circuits
were then investigated with a swap gate fidelity of $f_s \in \{0.999, 0.995, 0.99, 0.97, 0.95\}$.
Figure~\ref{fig:fid_boxplots}~shows the improvement in fidelity when using one-dimensional topology displacements in addition to noisy swap gates compared to a quantum circuit mapping with noisy swap gates only.
Using topology displacements in addition to swap gates lead to a maximal fidelity improvement of $29\%$ at a swap gate fidelity of $0.95$.
However, for every evaluated swap gate fidelity there existed quantum circuits whose fidelity did not improve, i.e. no swap gate could be replaced by a topology displacement.
At the largest evaluated swap gate fidelity $(0.999)$, the fidelity improved by $0.3\%$ on average.
For the lowest evaluated swap gate fidelity $(0.95)$, the quantum circuit fidelity improved by $15\%$ on average.
These results show that depending on the exact technology parameters, one-dimensional topology displacements can incur a significant improvement in fidelity or barely have an effect on the fidelity of the quantum circuit computation.

\section{Conclusion}\label{sec:conclusion}

\begin{spacing}{0.97} 
\normalsize
In this work a novel optimal quantum circuit method was developed that can exploit one-dimensional topology displacements available in near-term quantum architectures based on Rydberg atoms.
For the evaluated quantum circuits and technology parameters, the developed method incurred a quantum circuit depth reduction of up to 58\% and a fidelity improvement of up to 29\%.
We demonstrated that quantum circuit mapping can be improved through one-dimensional topology displacements if the swap gate fidelity is lower than $0.999$ or the swap gate runtime is not much lower than the topology displacement runtime.
The developed method can be used to map quantum circuits to near-term quantum architectures based on Rydberg atoms and provides technology parameters for experimentalists that can help extend the algorithmic opportunities of near-term Rydberg architectures.
\end{spacing}
\balance

\scriptsize

\bibliographystyle{IEEEtran}

\bibliography{rydberg_shifts}

\begin{thebibliography}{10}
\providecommand{\url}[1]{#1}
\csname url@samestyle\endcsname
\providecommand{\newblock}{\relax}
\providecommand{\bibinfo}[2]{#2}
\providecommand{\BIBentrySTDinterwordspacing}{\spaceskip=0pt\relax}
\providecommand{\BIBentryALTinterwordstretchfactor}{4}
\providecommand{\BIBentryALTinterwordspacing}{\spaceskip=\fontdimen2\font plus
\BIBentryALTinterwordstretchfactor\fontdimen3\font minus
  \fontdimen4\font\relax}
\providecommand{\BIBforeignlanguage}[2]{{%
\expandafter\ifx\csname l@#1\endcsname\relax
\typeout{** WARNING: IEEEtran.bst: No hyphenation pattern has been}%
\typeout{** loaded for the language `#1'. Using the pattern for}%
\typeout{** the default language instead.}%
\else
\language=\csname l@#1\endcsname
\fi
#2}}
\providecommand{\BIBdecl}{\relax}
\BIBdecl

\bibitem{drugdesign}
Y.~Cao, J.~Romero, and A.~Aspuru-Guzik, ``Potential of quantum computing for
  drug discovery,'' \emph{IBM Journal of Research and Development}, vol.~62,
  no.~6, pp. 6:1--6:20, 2018.

\bibitem{microsoft_nitrogen_fixation}
M.~Reiher, N.~Wiebe, K.~M. Svore, D.~Wecker, and M.~Troyer, ``Elucidating
  reaction mechanisms on quantum computers,'' \emph{Proc. Natl. Acad. Sci},
  vol. 114, no.~29, pp. 7555--7560, 2017.

\bibitem{materialscience}
B.~Bauer, S.~Bravyi, M.~Motta, and G.~Kin-Lic~Chan, ``Quantum algorithms for
  quantum chemistry and quantum materials science,'' \emph{Chemical Reviews},
  vol. 120, no.~22, pp. 12\,685--12\,717, 2020.

\bibitem{Shor}
P.~W. Shor, ``Algorithms for quantum computation: Discrete logarithms and
  factoring,'' in \emph{Proc. of the 35th Annual Symposium on Foundations of
  Computer Science}, 1994, p. 124–134.

\bibitem{Preskill2018}
J.~Preskill, ``Quantum computing in the {NISQ} era and beyond,''
  \emph{Quantum}, vol.~2, p.~79, 2018.

\bibitem{arsonisq}
S.~Brandhofer, S.~Devitt, and I.~Polian, ``{ArsoNISQ: Analyzing Quantum
  Algorithms on Near-Term Architectures},'' in \emph{Proc. of the 26th IEEE
  European Test Symposium}, 2021, pp. 1--6.

\bibitem{ionq}
\BIBentryALTinterwordspacing
\emph{IONQ Quantum Systems}, IONQ, Accessed 28th May, 2021. [Online].
  Available: \url{https://ionq.com/technology}
\BIBentrySTDinterwordspacing

\bibitem{ion2}
B.~Lekitsch, S.~Weidt, A.~G. Fowler, K.~M{\o}lmer, S.~J. Devitt, C.~Wunderlich,
  and W.~K. Hensinger, ``Blueprint for a microwave trapped ion quantum
  computer,'' \emph{Sci. Adv.}, vol.~3, no.~2, 2017.

\bibitem{quantum_inspire}
\BIBentryALTinterwordspacing
\emph{Quantum Inspire Quantum Systems}, QuTech, Accessed 28th May, 2021.
  [Online]. Available: \url{https://www.quantum-inspire.com/backends/}
\BIBentrySTDinterwordspacing

\bibitem{honeywell}
J.~M. Pino, J.~M. Dreiling, C.~Figgatt, J.~P. Gaebler, S.~A. Moses, C.~Baldwin,
  M.~Foss-Feig, D.~Hayes, K.~Mayer, C.~Ryan-Anderson \emph{et~al.},
  ``Demonstration of the {QCCD} trapped-ion quantum computer architecture,''
  2020.

\bibitem{GoogleSupremacy}
F.~Arute \emph{et~al.}, ``Quantum supremacy using a programmable
  superconducting processor,'' \emph{Nature}, vol. 574, no. 7779, pp. 505--510,
  2019.

\bibitem{ibm}
\BIBentryALTinterwordspacing
\emph{{IBM} Quantum Systems}, {IBM}, Accessed 28th May, 2021. [Online].
  Available: \url{https://quantum-computing.ibm.com/docs/manage/backends/}
\BIBentrySTDinterwordspacing

\bibitem{rigetti}
\BIBentryALTinterwordspacing
\emph{Rigetti Quantum Systems}, Rigetti, Accessed 28th May, 2021. [Online].
  Available: \url{https://qcs.rigetti.com/dashboard}
\BIBentrySTDinterwordspacing

\bibitem{rydberg_pascal}
A.~Browaeys and T.~Lahaye, ``Many-body physics with individually controlled
  {Rydberg} atoms,'' \emph{Nature Physics}, vol.~16, no.~2, pp. 132--142, 2020.

\bibitem{nv2}
K.~Nemoto \emph{et~al.}, ``Photonic architecture for scalable quantum
  information processing in diamond,'' \emph{Phys. Rev. X}, vol.~4, no.~3, p.
  31022, 2014.

\bibitem{swap_overhead}
A.~Cowtan, S.~Dilkes, R.~Duncan, A.~Krajenbrink, W.~Simmons, and S.~Sivarajah,
  ``On the qubit routing problem,'' 2019.

\bibitem{mapping_quantum_teleportation}
S.~Hillmich, A.~Zulehner, and R.~Wille, ``Exploiting quantum teleportation in
  quantum circuit mapping,'' in \emph{26th Asia and South Pacific Design
  Automation Conference}, 2021, pp. 792--797.

\bibitem{rydberg_high_fidelity_two}
I.~S. Madjarov, J.~P. Covey, A.~L. Shaw, J.~Choi, A.~Kale, A.~Cooper,
  H.~Pichler, V.~Schkolnik, J.~R. Williams, and M.~Endres, ``High-fidelity
  entanglement and detection of alkaline-earth {Rydberg} atoms,'' \emph{Nature
  Physics}, vol.~16, no.~8, pp. 857--861, 2020.

\bibitem{rydberg_multi_qubit_gate}
H.~Levine, A.~Keesling, G.~Semeghini, A.~Omran, T.~T. Wang, S.~Ebadi,
  H.~Bernien, M.~Greiner, V.~Vuleti\ifmmode~\acute{c}\else \'{c}\fi{},
  H.~Pichler, and M.~D. Lukin, ``Parallel implementation of high-fidelity
  multiqubit gates with neutral atoms,'' \emph{Phys. Rev. Lett.}, vol. 123, p.
  170503, 2019.

\bibitem{17}
F.~Nogrette, H.~Labuhn, S.~Ravets, D.~Barredo, L.~B\'eguin, A.~Vernier,
  T.~Lahaye, and A.~Browaeys, ``Single-atom trapping in holographic 2d arrays
  of microtraps with arbitrary geometries,'' \emph{Phys. Rev. X}, vol.~4, p.
  021034, 2014.

\bibitem{Barredo2016}
D.~Barredo, S.~De~L{\'e}s{\'e}leuc, V.~Lienhard, T.~Lahaye, and A.~Browaeys,
  ``An atom-by-atom assembler of defect-free arbitrary two-dimensional atomic
  arrays,'' \emph{Science}, vol. 354, no. 6315, pp. 1021--1023, 2016.

\bibitem{rydberg_optical_tweezers2}
M.~Endres, H.~Bernien, A.~Keesling, H.~Levine, E.~R. Anschuetz, A.~Krajenbrink,
  C.~Senko, V.~Vuletic, M.~Greiner, and M.~D. Lukin, ``Atom-by-atom assembly of
  defect-free one-dimensional cold atom arrays,'' \emph{Science}, vol. 354, no.
  6315, pp. 1024--1027, 2016.

\bibitem{QRydDemo}
\BIBentryALTinterwordspacing
\emph{{QRydDemo}}, {QRydDemo}, Accessed 28th May, 2021. [Online]. Available:
  \url{https://www.project.uni-stuttgart.de/qryddemo/}
\BIBentrySTDinterwordspacing

\bibitem{QUERA}
\BIBentryALTinterwordspacing
\emph{{QUERA}}, {QUERA}, Accessed 28th May, 2021. [Online]. Available:
  \url{https://www.quera-computing.com}
\BIBentrySTDinterwordspacing

\bibitem{Pasqal}
\BIBentryALTinterwordspacing
\emph{{Pasqal}}, {Pasqal}, Accessed 28th May, 2021. [Online]. Available:
  \url{https://pasqal.io}
\BIBentrySTDinterwordspacing

\bibitem{qc10th}
M.~A. Nielsen and I.~Chuang, \emph{Quantum computation and quantum
  information}.\hskip 1em plus 0.5em minus 0.4em\relax American Association of
  Physics Teachers, 2002.

\bibitem{k_brown}
K.~R. Brown, J.~Kim, and C.~Monroe, ``Co-designing a scalable quantum computer
  with trapped atomic ions,'' \emph{npj Quantum Information}, vol.~2, no.~1,
  pp. 1--10, 2016.

\bibitem{vts_special_session}
S.~Brandhofer, S.~Devitt, T.~Wellens, and I.~Polian, ``Special session: Noisy
  intermediate-scale quantum {(NISQ)} computers---how they work, how they fail,
  how to test them?'' in \emph{Proc. of the 39th IEEE VLSI Test Symposium},
  2021, pp. 1--6.

\bibitem{first_rydberg_gate_theory}
D.~Jaksch, J.~I. Cirac, P.~Zoller, S.~L. Rolston, R.~C\^ot\'e, and M.~D. Lukin,
  ``Fast quantum gates for neutral atoms,'' \emph{Phys. Rev. Lett.}, vol.~85,
  pp. 2208--2211, 2000.

\bibitem{first_rydberg_gate_experiment}
A.~Ga{\"e}tan, Y.~Miroshnychenko, T.~Wilk, A.~Chotia, M.~Viteau, D.~Comparat,
  P.~Pillet, A.~Browaeys, and P.~Grangier, ``Observation of collective
  excitation of two individual atoms in the {Rydberg} blockade regime,''
  \emph{Nature Physics}, vol.~5, no.~2, pp. 115--118, 2009.

\bibitem{first_rydberg_gate_experiment2}
E.~Urban, T.~A. Johnson, T.~Henage, L.~Isenhower, D.~Yavuz, T.~Walker, and
  M.~Saffman, ``Observation of rydberg blockade between two atoms,''
  \emph{Nature Physics}, vol.~5, no.~2, pp. 110--114, 2009.

\bibitem{graham2019}
T.~M. Graham, M.~Kwon, B.~Grinkemeyer, Z.~Marra, X.~Jiang, M.~T. Lichtman,
  Y.~Sun, M.~Ebert, and M.~Saffman, ``Rydberg-mediated entanglement in a
  two-dimensional neutral atom qubit array,'' \emph{Phys. Rev. Lett.}, vol.
  123, p. 230501, 2019.

\bibitem{wille_ion_trap}
R.~Wille, O.~Keszocze, and N.~Mohammadzadeh, ``Exact physical design of quantum
  circuits for ion-trap-based quantum architectures,'' in \emph{24th Design,
  Automation and Test Europe Conference}, 2021.

\bibitem{photon}
S.~Barz, ``Quantum computing with photons: Introduction to the circuit model,
  the one-way quantum computer, and the fundamental principles of photonic
  experiments,'' \emph{Phys B-At Mol Opt}, vol.~48, no.~8, p. 083001, 2015.

\bibitem{integrated_optics_topology_change}
R.~G. Hunsperger, \emph{Integrated optics}.\hskip 1em plus 0.5em minus
  0.4em\relax Springer, 1995.

\bibitem{swap_gates}
M.~Y. Siraichi, V.~F.~d. Santos, S.~Collange, and F.~M.~Q. Pereira, ``Qubit
  allocation,'' in \emph{Proc. of the International Symposium on Code
  Generation and Optimization}, 2018, pp. 113--125.

\bibitem{wille2014optimal}
R.~Wille, A.~Lye, and R.~Drechsler, ``Optimal swap gate insertion for nearest
  neighbor quantum circuits,'' in \emph{19th Asia and South Pacific Design
  Automation Conference}, 2014, pp. 489--494.

\bibitem{jku2}
A.~Zulehner, A.~Paler, and R.~Wille, ``An efficient methodology for mapping
  quantum circuits to the {IBM} {QX} architectures,'' \emph{IEEE Transactions
  on CAD}, vol.~38, no.~7, pp. 1226--1236, 2018.

\bibitem{wille_minh}
R.~Wille, L.~Burgholzer, and A.~Zulehner, ``Mapping quantum circuits to {IBM
  QX} architectures using the minimal number of {SWAP} and {H} operations,'' in
  \emph{56th Design Automation Conference}, 2019, pp. 1--6.

\bibitem{sabre}
G.~Li, Y.~Ding, and Y.~Xie, ``Tackling the qubit mapping problem for nisq-era
  quantum devices,'' in \emph{Proceedings of the 24th International Conference
  on Architectural Support for Programming Languages and Operating Systems},
  2019, pp. 1001--1014.

\bibitem{crosstalk_mapping}
P.~Murali, D.~C. McKay, M.~Martonosi, and A.~Javadi-Abhari, ``Software
  mitigation of crosstalk on noisy intermediate-scale quantum computers,'' in
  \emph{Proceedings of the 25th International Conference on Architectural
  Support for Programming Languages and Operating Systems}, 2020, pp.
  1001--1016.

\bibitem{variability_mapping}
S.~S. Tannu and M.~K. Qureshi, ``Not all qubits are created equal: a case for
  variability-aware policies for {NISQ}-era quantum computers,'' in
  \emph{Proceedings of the 24th International Conference on Architectural
  Support for Programming Languages and Operating Systems}, 2019, pp. 987--999.

\bibitem{OLSQ}
B.~Tan and J.~Cong, ``Optimal layout synthesis for quantum computing,'' in
  \emph{Proc. of the 39th International Conference on Computer-Aided Design},
  New York, NY, USA, 2020.

\bibitem{variability_mapping2}
S.~Nishio, Y.~Pan, T.~Satoh, H.~Amano, and R.~V. Meter, ``Extracting success
  from {IBM}’s 20-qubit machines using error-aware compilation,'' \emph{ACM
  Journal on Emerging Technologies in Computing Systems}, vol.~16, no.~3, pp.
  1--25, 2020.

\bibitem{ion_trap_mapping}
N.~Mohammadzadeh, ``Physical design of quantum circuits in ion trap
  technology--a survey,'' \emph{Microelectronics journal}, vol.~55, pp.
  116--133, 2016.

\bibitem{jku}
R.~Wille, D.~Gro{\ss}e, L.~Teuber, G.~W. Dueck, and R.~Drechsler, ``{RevLib}:
  An online resource for reversible functions and reversible circuits,'' in
  \emph{{Int'l Symp. on Multi-Valued Logic}}, 2008, pp. 220--225.

\bibitem{bv}
E.~Bernstein and U.~Vazirani, ``Quantum complexity theory,'' \emph{{SIAM}
  Journal on Computing}, vol.~26, no.~5, pp. 1411--1473, 1997.

\bibitem{qft}
D.~Coppersmith, ``An approximate {F}ourier transform useful in quantum
  factoring,'' 2002.

\bibitem{z3solver}
L.~De~Moura and N.~Bj{\o}rner, ``Z3: An efficient {SMT} solver,'' in
  \emph{International conference on Tools and Algorithms for the Construction
  and Analysis of Systems}.\hskip 1em plus 0.5em minus 0.4em\relax Springer,
  2008, pp. 337--340.

\end{thebibliography}

\end{document}